\documentclass[10pt,aip,cha,
 				twocolumn,reprint,
 				superscriptaddress,numerical]{revtex4-1}
\usepackage{graphicx} 
\usepackage{epstopdf}

\usepackage{dcolumn}   

\usepackage[utf8]{inputenc}
\usepackage[english] {babel}
\usepackage{graphicx}
\usepackage{natbib}
\usepackage{url}
\usepackage{amsfonts,amstext,amsmath}
\usepackage{bbm}
\usepackage{bm}
\usepackage{amssymb}
\usepackage{amsmath}
\usepackage{amsthm}
\usepackage{bbold}
\usepackage{fancybox,xcolor}
\usepackage{hyperref} 

\newcommand{\diag}{\ensuremath{\textrm{diag}}}

\newcommand{\tpeak}{\ensuremath{t^\mathrm{peak}}}

\newcommand{\be}{\begin{equation}}
\newcommand{\ee}{\end{equation}}
\newcommand{\bea}{\begin{eqnarray}}
\newcommand{\eea}{\end{eqnarray}}

\definecolor{shadowcolor}{RGB}{0, 0, 102}

\DeclareMathOperator*{\argmin}{arg\,min}

\theoremstyle{plain}

\newtheorem{lem}{Lemma}

\newtheorem{dfn}{Definition}
\newtheorem*{prf*}{Proof}

\begin{document}

\title{Quantifying Transient Spreading Dynamics on Networks}

\author{Justine Wolter}
\affiliation{Network Dynamics, Max Planck Institute for Dynamics and Self-Organization (MPIDS), 37077 G\"ottingen, Germany}
%
\author{Benedict L\"{u}nsmann}		
\affiliation{Max Planck Institute for the Physics of Complex Systems (MPIPKS), 01187 Dresden, Germany}
\author{Xiaozhu Zhang}
\affiliation{Network Dynamics, Max Planck Institute for Dynamics and Self-Organization (MPIDS), 37077 G\"ottingen, Germany}
\author{Malte Schr\"oder}
\affiliation{Network Dynamics, Max Planck Institute for Dynamics and Self-Organization (MPIDS), 37077 G\"ottingen, Germany}
\author{Marc Timme} 
\affiliation{Chair for Network Dynamics, Institute for Theoretical Physics and Center for Advancing Electronics Dresden (cfaed), 01069 Dresden, Germany}
\affiliation{Network Dynamics, Max Planck Institute for Dynamics and Self-Organization (MPIDS), 37077 G\"ottingen, Germany}
\affiliation{Bernstein Center for Computational Neuroscience (BCCN), 37077 G\"ottingen, Germany}
\affiliation{Department of Physics, Technical University of Darmstadt, 64289 Darmstadt, Germany}
\affiliation{ETH Z\"urich Risk Center, 8092 Z\"urich, Switzerland} 
 
\begin{abstract}
Spreading phenomena on networks are essential for the collective dynamics of various natural and technological systems, from information spreading in gene regulatory networks to neural circuits or from 
epidemics to supply networks experiencing perturbations. Still, how local disturbances spread across networks is not yet quantitatively understood.
Here we analyze generic spreading dynamics in deterministic network dynamical systems close to a given operating point. Standard dynamical systems' theory does not explicitly provide measures for arrival times and amplitudes of a transient, spreading signal because it focuses on invariant sets, invariant measures and other quantities less relevant for transient behavior. We here change the perspective and introduce effective expectation values for deterministic dynamics to work out a theory explicitly quantifying when and how strongly a perturbation initiated at one unit of a network impacts any other. The theory provides explicit timing and amplitude information as a function of the relative position of initially perturbed and responding unit as well as on the entire network topology.
\end{abstract}

\maketitle

\begin{quotation}
Networked systems characterize a large number of natural and man-made systems. Transient spreading phenomena fundamentally underlie the dynamics of these systems: an outbreak of a disease at one place may spread through a human mobility network on continental scales and a load shedding of a single power plant impacts distant parts of the power grid. It thus constitutes a natural question when, how long and how strongly such a perturbation affects other units in the network. Interestingly, this question does not possess a simple answer in standard dynamical systems theory, which often neglects such transient dynamics. In this article we introduce and analyze intuitive measures for typical response times and magnitudes via effective expectation values, interpreting the activity of each unit in the network as a probability density over time. We derive simple analytical expressions for these measures in linear dynamical systems. Across model systems, this makes it possible to analytically quantify transient spreading dynamics as a function of the network's interaction topology. 
\end{quotation}

\section{Introduction}
Many collective transient phenomena are initiated by perturbing some simple base state, for instance, a fixed (operating) point in a deterministic dynamical system or a stationary probability distribution of a Markov chain. For network dynamical systems, such initial perturbations often affect only a single unit and are thus local in the topology of the network. Examples range from the start of an epidemic in a population of susceptible agents (natural or artificial) \cite{brockmann13_spreading, Iannelli17_effectiveDistances} to the failure of a single infrastructure in a supply network \cite{witthaut16_critical, manik17_network_susceptibilities, kettmann16_acgrids, Menck14_powergrids, Xiaozhu17_networkPatterns}. If a single unit's variable is initially perturbed from a given fixed point value, other units in the network will be transiently affected by such a perturbation, with relevant consequences only at some later time. Natural questions thus include `At which time does a transient signal reach a given unit?' and `How strongly does the signal affect that unit?'

Despite the growing interest in spreading and  propagation processes, non-trivial waves, and other transient phenomena \cite{newman02_epidemics, hufnagel04_forecast, brockmann06_travel, brockmann13_spreading, timme04_speed_limits, timme06_synch_speed, witthaut15_cascadingfailures, witthaut13_singlelinkaddition, degueldre16_tsunami, Xiaozhu17_networkPatterns, kittel17_quantifying_transients}, there is no general answer to these questions. For certain stochastic systems, there is recent mathematical progress on quantifying first arrival and routing times in stochastic systems \cite{Iannelli17_effectiveDistances, braunstein03_optimalpaths, gautreau07_arrivaltimesDisease, roosta82_reliableRouting}. For simple deterministic dynamical systems major questions remain open, mainly because existing mathematical theory of such systems is restricted to mostly two relevant classes of general statements: one about long term behavior, characterized by different types of invariant sets such as attractors in dissipative systems, and a second about statistical properties such as those captured by invariant measures in chaotic and stochastic dynamical systems. These two classes of statements both do not explicitly capture transient phenomena.

Mathematically, for instance, it becomes impossible to provide an explicit formula for the time $\tpeak_i$ of the maximum magnitude of the transient signal at a given unit $i$ as soon as the network topology becomes non-trivial. Even for linear dynamics this impossibility persists because the task is equivalent to solving a transcendental equation. The same problem transfers to the signal amplitude at that time such that quantifying arrival times and perturbation impact in network dynamical systems constitutes an open challenge.

Here we propose an alternative perspective to characterize transient spreading dynamics in network dynamical systems. We do not attempt to approximate peak positions of maxima of the units' variables in time and the respective peak heights. For a class of local dynamics close to given operating (fixed) points, we instead interpret the units' state trajectories $x_i(t)$ as if they were measure densities and, once suitably normalized, yield effective probability densities. From these we analytically derive integral expressions (effective expectation values representing zeroth, first and second moments) that define typical times and response magnitudes as an explicit function of the matrix determining the network topology.

\section{Network Dynamical System and Problem Setting}
Consider a network dynamical system
\be
\frac{\mathrm{d}\mathbf{y}}{\mathrm{d}t}=\mathbf{F}(\mathbf{y})
\label{eqn:nonlinearsystem_general}
\ee
of $N$ coupled units whose collective dynamics is close to a stable operating point $\mathbf{y}^* \in \mathbb{R}^N$ where $\mathbf{F}(\mathbf{y^*})=\mathbf{0}$. The system's dynamics can then be specified in new difference variables $\mathbf{x}(t)=\mathbf{y}(t)-\mathbf{y}^*$
satisfying linear equations of the type
\be
\frac{\mathrm{d}\mathbf{x}}{\mathrm{d}t}=M\mathbf{x} \,,
\label{eqn:linearsystem_general}
\ee
where $\mathbf{x}(t)=\left(x_1(t),\ldots, x_N(t) \right)^\mathrm{T} \in \mathbb{R}^N$ defines the states $x_i(t)$ of the unit $i$ at time $t \in \mathbb{R}$ and $M = \mathrm{D}\mathbf{F}(y^*) \in \mathbb{R}^{N\times N}$ is a weighted matrix. We consider $M$ to have only non-negative off-diagonal elements with $M_{ij}=0$ if there is no direct interaction from unit $j$ to unit $i$. Wherever an element $M_{ij} > 0$ for $j\neq i$, unit $i$ is directly coupled to $j$. Such systems arise not only in network dynamical systems [Eq.~(\ref{eqn:nonlinearsystem_general})] of coupled units $i$ near fixed (operating) points, but also naturally occur in time-continuous master equations for probabilities $\mathbb{P}_i(t)\equiv x_i(t)$ of the system to be in state $i$ at time $t$ \cite{norris98_markovbook}.


For a specific example class that we use for illustration below, consider a strongly connected directed graph $G$ with weighted graph adjacency matrix $A$ with elements $A_{ii}=0$ and $A_{ij}\geq 0$ for $i\neq j$.
A graph is strongly connected if there are directed paths $k\rightarrow ...\rightarrow j$ from every unit $k$ to  every other $j$. If a graph has several disconnected, strongly connected components, we consider each strongly connected component separately.
If and only if the underlying graph has a directed edge from unit $j$ to unit $i$, we have  $A_{ij} = \alpha_{ij} > 0$.
The associated graph Laplacian is $L=D-A$ where the diagonal matrix $D$ has entries
\be
D_{ii}=\sum_{j=1}^{N} A_{ij}
\ee
for $i \in \{1,\ldots,N\}$ and $D_{ij}=0$ for $i\neq j$. The matrix $M=-L-\diag_{i\in \{1,\ldots,N\}}\left(\beta_i\right)$ then describes internal dynamics of the individual units ($\beta_i$) as well as the coupling between the units ($\alpha_{ij}$).
\begin{dfn}\label{dfn:linearsystem}
We consider the linear system of coupled units described by
\bea
\label{eqn:linearsystem}
\frac{\mathrm{d}x_i}{\mathrm{d}t} &=& \left( M \mathbf{x} \right)_i = -\beta_i x_i + \sum_{j=1}^{N} \alpha_{ij} \left(x_j - x_i \right) \,, \\ 
x_i(0) &\ge& 0	\notag\,,
\eea
where at least one node $k \in \left\{1, \dots, N\right\}$ is initially perturbed $x_k(0) > 0$. Further, we assume $\alpha_{ij} \ge 0$ and $\beta_i > 0$, such that the matrix $M$ is irreducible, negative diagonally dominant and consequently all eigenvalues have negative real part.
\end{dfn}
This definition assures that $\mathbf{x}^* = (0,0, \dots, 0)^T$ is a stable fixed point and the system describes a strongly connected network where a perturbation of one unit can reach any other.

How does a perturbation applied at some unit $k$ spread across the network? When and how strongly do other units $i$ respond to the initial perturbation? How do these responses depend on the relative locations of the units and the features of the network topology? 
For the linear dynamical system (Def.~\ref{dfn:linearsystem}) with a single perturbed node 
\be
\mathbf{x}_0:=\mathbf{x}(0)=(0,0,\ldots,\underbrace{1}_{x_k(0)}, 0,\ldots,0)
\label{eqn:initialcondition}
\ee
the complete time-dependent trajectory 
\be
x_i(t) = \left[\exp(M t) \mathbf{x}_0\right]_i = \left[ \exp(M t) \right]_{ik}
\label{eqn:solution}
\ee
 is known analytically [here  $\exp(\cdot)=e^{\cdot}$ is the matrix exponential]. Yet, a number of key obstacles hinder immediate answers, as we will see below. In the current article, we contribute to exactly specifying and analytically answering the open questions raised above by changing the perspective about how to address them.

\begin{figure*}
\includegraphics[width=140mm]{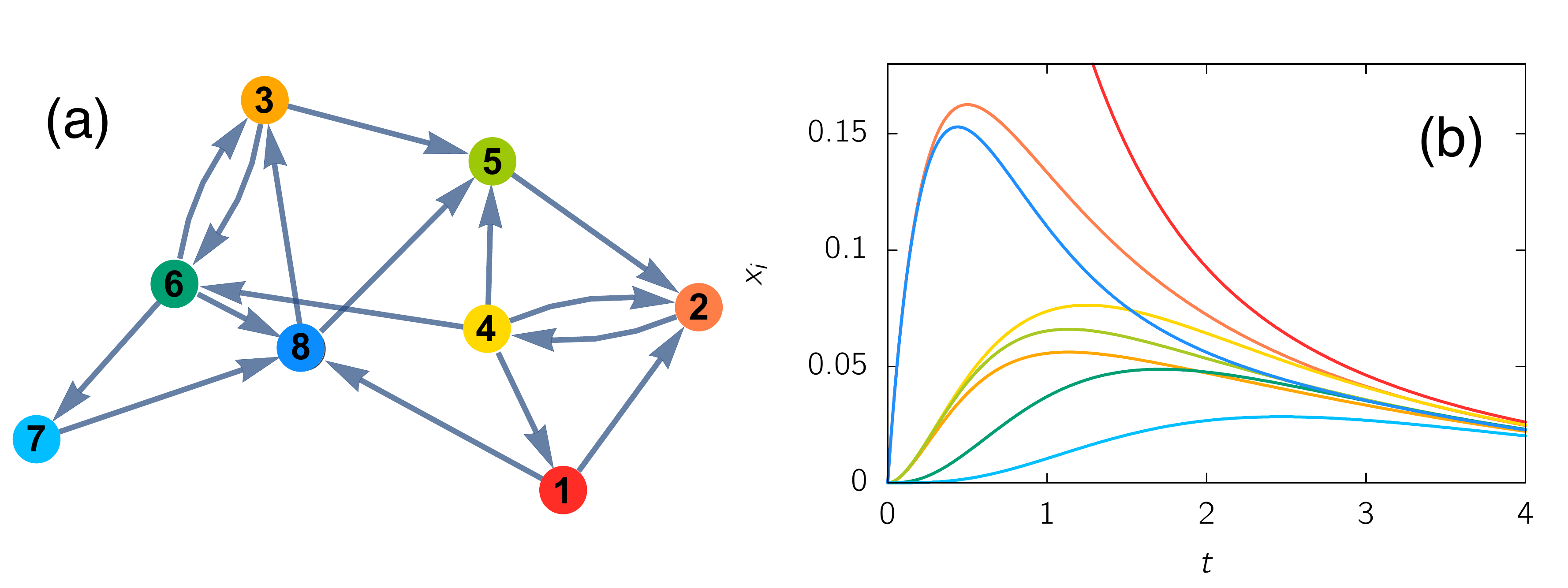}
 \caption{ 
 \label{fig:illustration}
 (color online)
\textbf{Dynamics of a transient perturbation spreading in a network of coupled units.}
(a) Network with $N=8$ units and $\left|E\right|=16$ links describing directed interactions with $\alpha_{ij} = 1$ and $\beta_i = \beta = 0.5$ (Def.~\ref{dfn:linearsystem}). (b) Response for an initial perturbation at unit $k=1$, i.e., with initial condition $x_i(0)=\delta_{i,1}$. The response of each node is colored according to the node color in panel (a). Whereas the activity of the initially perturbed unit $k=1$ (red curve) decays exponentially, the activity dynamics of all other units are non-monotonic and generically exhibit (at least) one maximum.
}
\end{figure*}

\section{Trancendental equations determine standard response times}
\label{sec:standard}

Direct numerical simulations across a range of random and regular network topologies (Figure \ref{fig:illustration}) suggest that the units respond to an initial perturbation in a characteristic way: we observe that, as expected, all but the initially perturbed units' state variables grow from zero to positive values, then decay to zero exponentially and thus exhibit (at least) one maximum in between. 
In this article, we focus on the question of how to characterize and quantify such transient dynamics, in particular by determining the unit-dependent response times and response strengths (Figure \ref{fig:illustration}b).

One natural characteristic measure for the response time that may be interpreted as the time of signal arrival at unit $i$ is the peak time $\tpeak_{i}$ where
\be
\left.\frac{\mathrm{d}x_i}{\mathrm{d}t}\right|_{\tpeak_{i}}=0 \quad \mathrm{and} \quad \left.\frac{\mathrm{d}^2x_i}{\mathrm{d}t^2}\right|_{\tpeak_{i}}<0
\label{eqn:maximumCondition1}
\ee 
and the activity amplitude $x_i(\tpeak_{i})$ at that time. In the general case when there may be multiple extrema satisfying Eq.~(\ref{eqn:maximumCondition1}), we define $\tpeak_{i}$ as the time of the first maximum. Despite knowing the complete analytic solution [Eq.~(\ref{eqn:solution})], an analytic expression for these times does not exist. In fact, the equations
\be
\sum_{j=1}^N M_{ij} \left( e^{M \tpeak_{i}}\mathbf{x}_0\right)_{j} \stackrel{ \textrm{!}}{=} 0
\label{eqn:maximumCondition2}
\ee 
that determine the maximum times $\tpeak_{i}$ contain differently weighted sums of different exponentially decaying functions and are not only implicit but also typically transcendental. Only under strong conditions, for instance if the system is very sparse such that unit $i$ receives only one connection from one other unit in the network, Eq.~(\ref{eqn:maximumCondition2}) becomes analytically solvable for $\tpeak_{i}$.

A second candidate measure for a characteristic response time is the time $t^{(c)}_i$ until the activity at unit $i$ increased above a certain predefined constant $x_i(t^{(c)}_i)\geq c$.  In analogy to $\tpeak_{i}$ in Eq.~(\ref{eqn:maximumCondition1}), the time
\be
t^{(c)}_i=\argmin_{t>0} \left\{ \left( e^{M t} \mathbf{x}_0 \right)_{i} \geq c \right\}
\ee
is again given implicitly by a transcendental equation. Moreover, this time $t^{(c)}_i$ depends on an arbitrary parameter $c$ that is additionally introduced and, if chosen too large, $t^{(c)}_i$ may not even exist for some units $i$.

\section{Alternative perspective on response times}
\label{sec:alternative}
In the following, we propose an alternative perspective to characterize typical response times and response magnitudes in linear network dynamical systems of arbitrary interaction topology. Instead of attempting to approximate peak positions or threshold crossing times discussed above, we first show that the units' state trajectories $x_i(t)$ are positive for all times $t>0$. Normalizing them we can interpret the new quantity $\rho_i(t) \propto x_i(t)$ as a probability density and use analogues to expectation values such as $\left< t \right>_i:=\int_0^\infty t \rho_i(t) dt$ 
to define typical response times, response durations and response magnitudes.

To be able to exploit the analogy to probability densities, we first establish positivity.

\begin{lem}[All component dynamics are positive]
\label{lem:positivity}
The system given in Def.~\ref{dfn:linearsystem} has positive activities of all units for all positive times: the solution $\mathbf{x}(t)$ of Eq.~(\ref{eqn:linearsystem}) satisfies $x_i(t)>0$ for all $t\in (0,\infty)$ and all  $i \in \{1,\ldots,N\}$.
\end{lem}

\begin{prf*}
The solution dynamics [Eq.~(\ref{eqn:solution})] of unit $i$ are given by
\be
x_i(t) = \sum_{j=1}^N \left(e^{M t}\right)_{ij} x_j(0) .
\ee 
Define a matrix
\be
C := M + b I_N \,,
\ee
where $b>\max_i\{|M_{ii}|\}$ and $I_N \in \mathbb{R}^{N \times N}$ denotes the identity matrix. Then $C$ is an irreducible matrix with strictly positive diagonal entries $C_{ii} > 0$ and non-negative off-diagonal entries. Consequently, all entries $[C^{n}]_{ij} \ge 0$ for all $n \ge 0$ and there exists $n^* \in \mathbb{N}$ such that, for all $n \ge n^*$, $C^{n}$ is strictly positive, that means $[C^{n}]_{ij} > 0$ for all $i,j \in \{1,2 \dots N\}$ ($C$ is a primitive matrix). Consequently, the matrix exponential is also strictly positive, $\left( e^{Ct}\right)_{ij} = \sum_{n=0}^\infty t^n \left[C^n\right]_{ij} / n! > 0$, for all $i,j \in \{1,\ldots,N\}$ and all positive $t$. Thus, we have for all $t > 0$
\bea
x_i(t) &=& \sum_{j=1}^N \left(e^{M t}\right)_{ij} x_j(0) \notag\\
&=& \sum_{j=1}^N \left[e^{C t - b I_N t}\right]_{ij} x_j(0) \notag\\
&=& \sum_{j=1}^N e^{-b t} \left( e^{C t} \right)_{ij} x_j(0) > 0 
\eea
for all $i,k \in \{1,\ldots,N\}$. 
\qed
\end{prf*}

Understanding that the response of a unit is always positive, it is natural to define the \textit{total response strength} $Z_i$.
\setcounter{dfn}{1}
\begin{dfn}[Total response strength]
	The total response strength $Z_i$ of a unit $i$ is given by
	\be
	\boxed{
		Z_i := \int_0^{\infty} x_i(t) dt \,.
		}
	\label{eqn:partitionFunction}
	\ee
\end{dfn}

Since the analytical solution for $x_i(t)$ is known [Eq.~(\ref{eqn:solution})], we can express $Z_i$ in terms of the matrix $M$ defining the system in Def.~\ref{dfn:linearsystem}).
\begin{lem}[Total response strength]
\label{lem:total_response}
	The total response strength $Z_i$ of a unit $i$ is given by
	\be
	\boxed{
		Z_i = - \left(M^{-1} \mathbf{x}_0 \right)_i \,.
		}
	\label{eqn:partitionFunction_matrix}
	\ee
\end{lem}

\begin{prf*}
\begin{align}
Z_i &=\int_0^{\infty} x_i(t) dt\notag\\
&\overset{(\ref{eqn:solution})}=\left[\int_0^{\infty} \exp(M t)  \mathbf{x}_0\right]_i\notag\\
&=\left[M^{-1}\exp(M t)\mathbf{x}_0\bigr\rvert_0^{\infty} \right]_i \notag\\
&=- \left(M^{-1} \mathbf{x}_0 \right)_i \notag \,.
\end{align}

Alternatively, we can simply integrate the differential equation [Eq.~(\ref{eqn:linearsystem})], see Appendix~\ref{app:proof}.\qed
\end{prf*}

In particular, for initial perturbation of a single unit $k$ [Eq.~(\ref{eqn:initialcondition})], we obtain
\be
\label{eqn:partitionfunctionSpecific}
Z_i = - (M^{-1})_{ik} \,.
\ee

Given this definition of the response strength and the positivity of the units' response dynamics $x_i(t)$ established in lemma~{\ref{lem:positivity}, we interpret the \emph{response density}
\be
\boxed{
\rho_i(t):=\frac{x_i(t)}{Z_i}
}
\label{eqn:probability}
\ee
as a probability density.
Note that standard response characteristics, such as the peak response time $\tpeak_{i}$, correspond to standard characteristics of a probability distribution, such as the mode of distribution. We follow this similarity and interpret also the expectation values as typical response characteristics, all of which are illustrated and summarized in Fig.~\ref{fig:quantities}.

\begin{figure}
\centering
\includegraphics[width=0.45\textwidth]{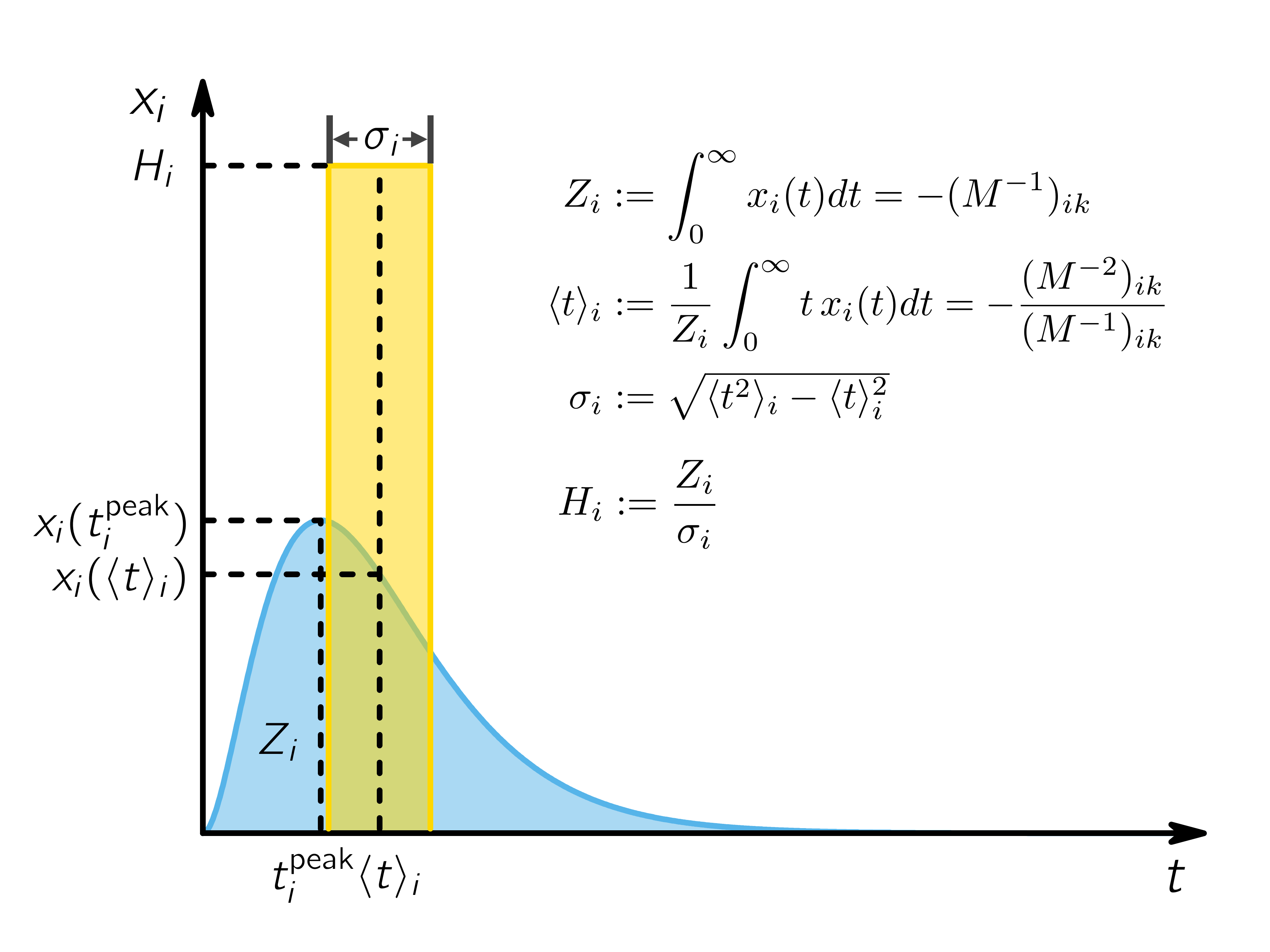}
\caption{ (color online)
\textbf{Quantifying the response to perturbations.}
Illustration of the typical response characteristics derived from the interpretation of the response $x_i(t)$ to a perturbation at unit $k$ as a probability density over $t$ [Eq.~(\ref{eqn:probability})]. The effective expectation value $\left<t\right>_{i}$ describes the typical response time and the standard deviation $\sigma_i$ describes the typical response duration. Assuming a response with fixed magnitude for the duration $\sigma_i$ with total impact $Z_i$, $H_i$ then describes the typical response magnitude. Standard measures such as the peak response time $\tpeak_{i}$ and the response amplitude $x_i(\tpeak_{i})$ correspond to the mode of the ``distribution''.
}
\label{fig:quantities}
\end{figure}

We interpret the effective expectation value of $t$ with respect to $\rho_i(t)$ as the \emph{typical response time} of unit $i$.

\begin{dfn}[Typical response time]
	The typical response time $\left<t\right>_i$ of a unit $i$ is given by
	\be
	\boxed{
		\left<t\right>_i := \int_0^{\infty}  t \rho_i(t) dt
		}
	\label{eqn:meantime}
	\ee
\end{dfn}

\begin{lem}[Typical response time]
\label{lem:typical_time}
	The typical response time $\left<t\right>_i$ of a unit $i$ is given by
	\be
	\boxed{
		\left<t\right>_i = -\frac{(M^{-2} \mathbf{x}_0)_i}{(M^{-1} \mathbf{x}_0)_i} 
		}
	\label{eqn:meantime_matrix}
	\ee
\end{lem}

\begin{prf*}
We first calculate
\bea
\int_0^{\infty} t x_i(t) dt &=& \int_0^{\infty} t \left[\exp(M t) \mathbf{x}_0\right]_i dt \notag\\
&=& \Big[ M^{-1}  t \exp(Mt) \mathbf{x}_0  \bigr\rvert_0^{\infty} \notag\\
&\quad&\quad - \int_0^{\infty} M^{-1} \exp(Mt) \mathbf{x}_0~dt \Big]_i\notag\\
&=& \left[ -M^{-2} \exp(Mt) \mathbf{x}_0  \bigr\rvert_0^{\infty} \right]_i\notag\\
&=&\left(M^{-2} \mathbf{x}_0 \right)_i\notag \,.
\eea
Together with lemma \ref{lem:total_response} and the definition of $\rho_i(t)$ [Eq. (\ref{eqn:probability})] we arrive at the result. \qed
\end{prf*}

In particular, for initial perturbation of a single unit [Eq.~(\ref{eqn:initialcondition})] we obtain
\be
\label{eqn:meantime_specific}
\left<t\right>_i = - \frac{(M^{-2})_{ik}}{(M^{-1})_{ik}} \,.
\ee

This provides information on the time when a perturbation affects a unit. In order to also obtain a measure describing how strongly this unit is affected, we similarly interpret the standard deviation as the \emph{typical response duration}.
\begin{dfn}[Typical response duration]
	The typical response duration $\sigma_i$ of a unit $i$ is given by
	\be
	\boxed{
		\sigma_i := \sqrt{\int_0^{\infty}  (t-\left<t\right>_i)^2 \rho_i(t) dt}
		}
	\label{eqn:duration}
	\ee
\end{dfn}

\begin{lem}[Typical response duration]
\label{lem:typical_duration}
	The typical response duration $\sigma_i$ of a unit $i$ is given by
	\be
	\boxed{
		\sigma_i = \sqrt{ \frac{2(M^{-3} \mathbf{x}_0)_i}{(M^{-1} \mathbf{x}_0)_i} - \left(\frac{(M^{-2} \mathbf{x}_0)_i}{(M^{-1} \mathbf{x}_0)_i}\right)^2 }
		}
	\label{eqn:duration_matrix}
	\ee
\end{lem}

\begin{prf*}
We first calculate
\bea
\int_0^\infty t^2  \exp(M t)  \mathbf{x}_0 dt &=&  t^2 M^{-1} \exp(Mt) \mathbf{x}_0 \bigr\rvert_0^{\infty} \notag\\
	&\quad&\quad - \int_0^{\infty} 2t M^{-1} \exp(Mt) \mathbf{x}_0 dt\notag\\
&=&-2 t M^{-2} \exp(Mt) \mathbf{x}_0  \bigr\rvert_0^{\infty} \notag\\
	&\quad&\quad + \int_0^{\infty} 2 M^{-2} \exp(Mt) \mathbf{x}_0 dt\notag\\
&=&+2 M^{-3} \exp(Mt) \mathbf{x}_0 \bigr\rvert_0^{\infty}\notag\\
&=& -2 M^{-3} \mathbf{x}_0 \,.
\eea
We thus obtain $\left<t^2\right>_i = \frac{-2 \left(M^{-3} \mathbf{x}_0\right)_i}{Z_i}$. The lemma then follows directly from $\left<\left(t - \left<t\right>_i\right)^2\right>_i = \left<t^2\right>_i-\left<t\right>_i^2$ and equations (\ref{eqn:partitionFunction_matrix}) and (\ref{eqn:meantime_matrix}). \qed
\end{prf*}

For the particular initial condition [Eq.~(\ref{eqn:initialcondition})] this becomes
\be
\label{eqn:duration_specific}
\sigma_i = \frac{\sqrt{2(M^{-3})_{ik} (M^{-1})_{ik}- (M^{-2}_{ik})^2}}{-(M^{-1})_{ik}} \,.
\ee

The definition of the \emph{typical response magnitude}, describing how strongly a unit is affected by the initial perturbation, then follows naturally as the quotient of the total strength $Z_i$ and the duration $\sigma_i$, illustrated in Fig.~\ref{fig:illustration}}.
\begin{dfn}[Typical response magnitude]
	The typical response magnitude $H_i$ of a unit $i$ is given by
	\be
	\boxed{
		H_i := Z_i / \sigma_i = \frac{((M^{-1} \mathbf{x}_0)_i)^2}{\sqrt{2(M^{-3} \mathbf{x}_0)_i (M^{-1} \mathbf{x}_0)_i- ((M^{-2} \mathbf{x}_0)_i)^2}}
		}
	\label{eqn:height}
	\ee
\end{dfn}
For the particular initial condition [Eq.~(\ref{eqn:initialcondition})] this becomes
\be
\label{eqn:height_specific}
H_i =  \frac{(M^{-1}_{ik})^2}{\sqrt{2(M^{-3})_{ik} (M^{-1})_{ik}- (M^{-2}_{ik})^2}} \,.
\ee

The above definitions Eq.~(\ref{eqn:partitionFunction}), (\ref{eqn:meantime}), and (\ref{eqn:height}) thus yield explicit analytically derived quantifiers for the total response strength, the typical response time and the typical response magnitude of unit $i$.
They hold for arbitrary strongly connected network topologies. We note again that a graph is strongly connected if there is a directed path from every unit to every other unit, not implying anything about the coupling strength between units. If there is no directed path from unit $k$ to $i$, the perturbation cannot reach this node and $x_i(t) = 0$ for all $t$. Consequently, we have a total impact of $Z_i = (M^{-1})_{ik} = 0$ such that the arrival time and the other measures are not defined.

\section{Illustrating examples}
\label{sec:examples}
\subsection{Directed homogeneous chains}

\begin{figure*}
\centering
\includegraphics[width=130mm]{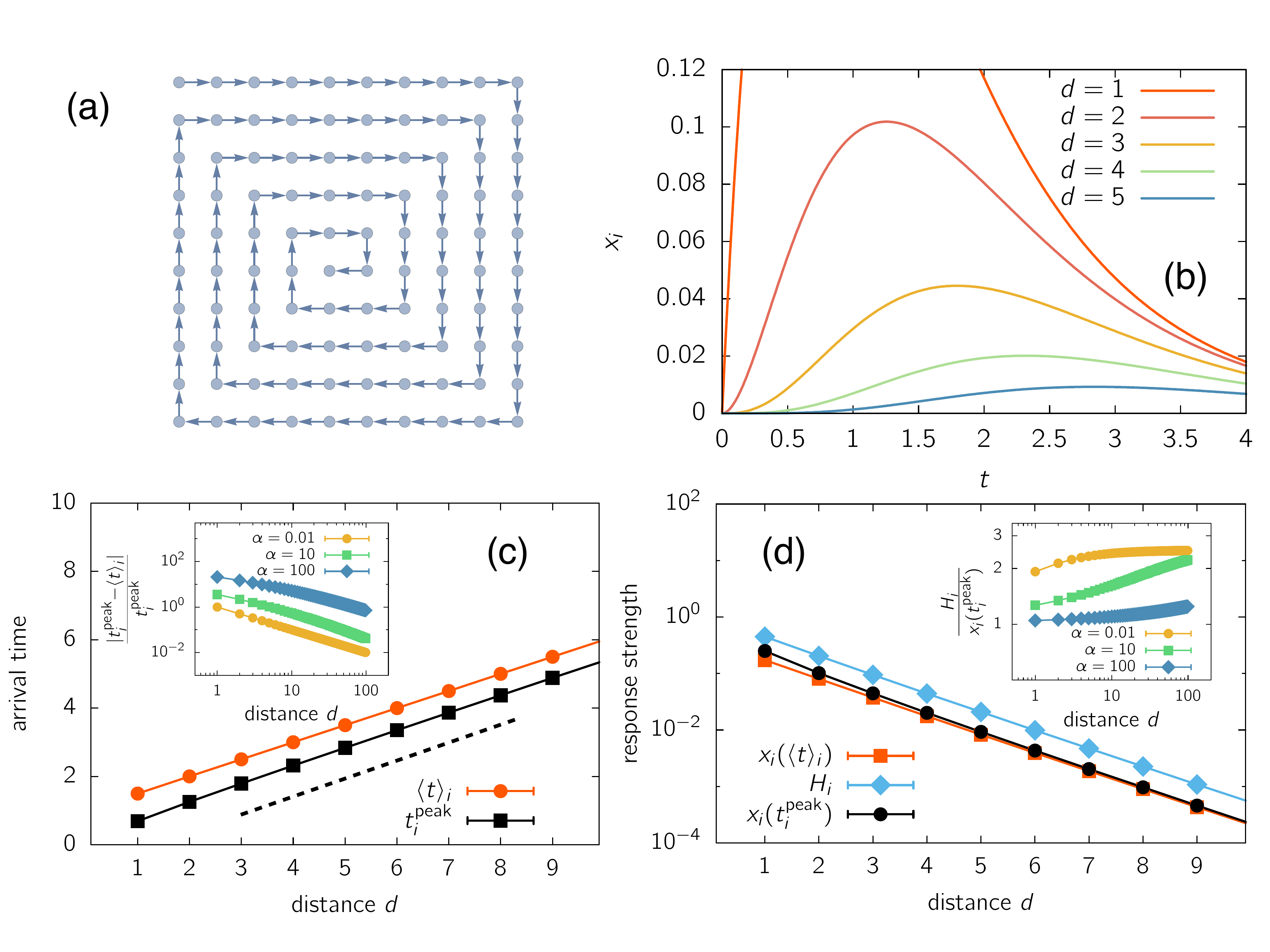}
\caption{ (color online)
\textbf{Perturbation spreading in a directed chain.}
Illustration of the measures introduced above, describing approximate time and impact of the perturbation. (a) Illustration of the network topology, a directed with $N=100$ nodes and $\beta = 1$. The node at the beginning of the chain was perturbed. (b) Response dynamics of the first 5 nodes in the chain. (c) Comparison of the typical response time $\left<t\right>_i$ and the peak response time $\tpeak_i$ for $\alpha = 1$. Both values scale identically with increasing distance from the perturbation and the absolute difference is almost constant. The perturbation spreads with a constant speed, as predicted by Eq.~(\ref{eqn:chain_speed}) and shown as the black dashed line. For large distances the approximation becomes more and more accurate, the relative distance decays for increasing $d$, independent of $\alpha$ (see inset). (d) Measurements of the strength of the perturbation given by the typical response magnitude $H_i$, the response at the typical response time $x_i\left(\left<t\right>_i\right)$ and the response amplitude $x_i\left(\tpeak_i\right)$. All values scale identically with increasing distance from the perturbation. The height $H_i$ overestimates $x_i\left(\tpeak_i\right)$ by a constant factor, when the distance is large compared to the coupling strength (see inset).}
\label{fig:directedchain}
\end{figure*}

\begin{figure*}
\centering
\includegraphics[width=130mm]{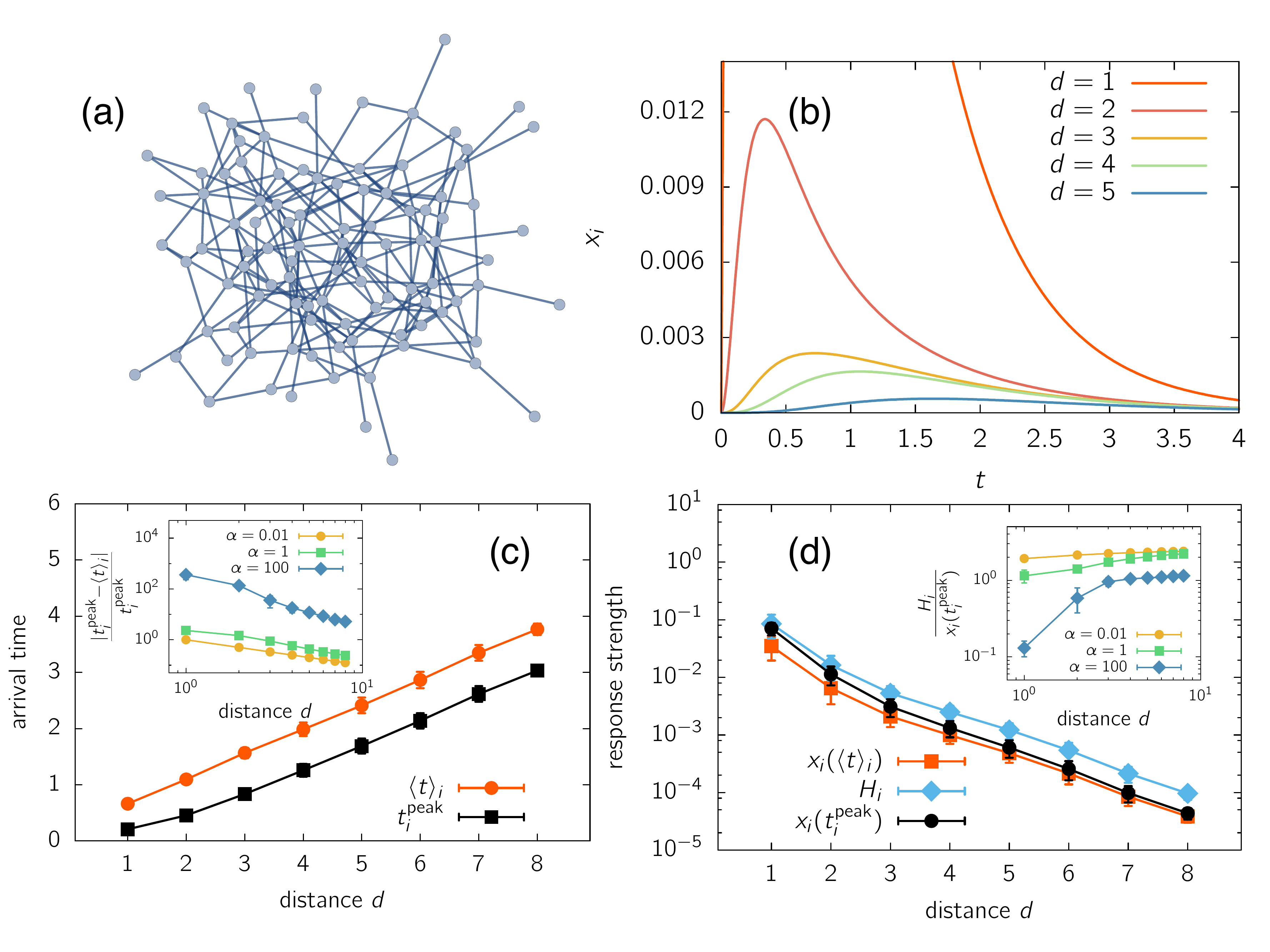}
\caption{ (color online)
\textbf{Consistent quantification of perturbation spreading in random networks.}
Illustration of the measures introduced above, describing approximate time and impact of the perturbation. (a) Illustration of the network topology, a connected random network with $N=100$ nodes, $M = 200$ and $\beta = 1$. All results are averaged over $10$ different realizations of the network topology and perturbation of all nodes. Error bars indicate the standard deviation across these 1000 realizations of transient spreading, 
indicating a high degree of consistency across network topologies and perturbation sites. (b) Example response dynamics of 5 nodes in the network with different distances to the initial perturbation. (c) Comparison of the typical response time $\left<t\right>_i$ and the peak response time $\tpeak_i$ for $\alpha = 1$. (d) Measurements of the strength of the perturbation given by the typical response magnitude $H$, the response at the typical response time $x_i\left(\left<t\right>_i\right)$ and the  response amplitude $x_i\left(\tpeak_i\right)$.}
\label{fig:random_network}
\end{figure*}

To illustrate these response quantifiers  we consider a basic example, a directed chain network which consist of $N$ units coupled only to one neighboring unit via a directed link with identical coupling strength $\alpha_{ij} \equiv \alpha$ [Fig.~\ref{fig:directedchain}(a)]. We assume for each unit $i$ identical internal dynamics $\beta_i \equiv \beta$. The dynamics of the directed homogeneous chain is then given by
\begin{align}
\dot{\mathbf{x}}&=\begin{pmatrix}
 -\beta& 0 &\dots & 0\\
\alpha & -(\beta+\alpha)&\dots&0\\ 
\vdots &\ddots &\ddots&\vdots\\
 0& \dots &\alpha & -(\beta+\alpha)
 \end{pmatrix}  \mathbf{x} \,.
\end{align}

We consider the initial condition $\mathbf{x}_0=(1,0,\ldots,0)$, perturbing the first unit $k=1$ in the chain. In this particular case, all quantities characterizing the response behavior can be written explicitly as functions of the parameters $\alpha$ and $\beta$ and the position of the node $i$. This allows us to gain intuition about how the measures proposed in the previous section quantify perturbation spreading in networks and to see how they compare to the standard measures.

The trajectory of each node $i$ can be solved analytically, which reads
\begin{equation}
x_i(t)=
\begin{cases}
e^{-\beta t} & \text{for }i=1\\
e^{-\beta t}  \left(1 - e^{-\alpha t}  \displaystyle\sum_{j=0}^{i-2} \dfrac{(\alpha t)^j}{j!} \right) &\text{for }i\geq 2 \, .\\
\end{cases}
\label{eqn:chain_solution}
\end{equation}
As discussed in Sec.~\ref{sec:standard}, the measures characterizing the network response latency, the signal arrival time $\tpeak_i$ and the activity amplitude $x_i(\tpeak_i)$, cannot be determined analytically via Eq.~(\ref{eqn:maximumCondition1}). 

Alternatively, we analytically quantify the network response times and response magnitudes from the probabilistic perspective we proposed in Sec.~\ref{sec:alternative}. Using Eqns.~(\ref{eqn:partitionFunction}), (\ref{eqn:probability}), (\ref{eqn:meantime}), (\ref{eqn:duration}), and (\ref{eqn:height}), we obtain the effective partition function $Z_i$, the effective probability density $\rho_i(t)$, the typical response time $\langle t \rangle_i$, the typical response duration $\sigma_i$, and the typical response magnitude $H_i$ as follows:
\begin{equation}
Z_i = \dfrac{1}{\beta}\left({\dfrac{\alpha}{\alpha +\beta}}\right)^{i-1} \,,\label{eqn:chain_partitionfunction}
\end{equation}
\begin{equation}
\rho_i(t) = \left({\frac{\alpha}{\alpha +\beta}}\right)^{1-i}\beta  e^{-\beta t}  \left(1 - e^{-\alpha t}  \sum\limits_{j=0}^{i-2} \dfrac{(\alpha t)^j}{j!} \right) \,,\label{eqn:chain_probability}
\end{equation}
\begin{equation}
\langle t \rangle_i = \dfrac{\alpha + i \beta}{\alpha \beta + {\beta}^2} \,,\label{eqn:chain_meantime}
\end{equation}
\begin{equation}
{\sigma}_{i} = \frac{\sqrt{{\alpha}^2+2\alpha \beta+ i {\beta}^2}}{(\alpha \beta + {\beta}^2)} \,,\label{eqn:chain_duration}
\end{equation}
\begin{equation}
H_i = \frac{(\alpha+\beta)}{\sqrt{{\alpha}^2+2\alpha \beta+i {\beta}^2}}\left(\frac{\alpha}{\alpha+\beta}\right)^{i-1}\,.\label{eqn:chain_height}
\end{equation}
For the detailed calculations see Appendix~\ref{sec:homogeneousChainApp}. Together with the graph distance $d \equiv d(i,1) = i - 1$ between node $i$ and the initially perturbed node $1$, these equations also provide an explicit dependence.

How do these novel quantities characterize the perturbation spreading in networks compared with the common measures, i.e., the first peak time $\tpeak_i$ and the activity amplitude $x_i(\tpeak_i)$? \\

Fig.~\ref{fig:directedchain}(b) shows the response of the first few nodes in the chain. The further from the source of the perturbation, the later the response occurs and the weaker it is. Both the peak response time $\tpeak_i$ as well as the typical response time $\left<t\right>_i$ show that the perturbation propagates through the chain with a constant speed [see Fig.~\ref{fig:directedchain}(c)]. Using Eq.~(\ref{eqn:chain_meantime}) we calculate the speed with respect to the distance $d$ from the origin of the perturbation as 
\begin{align}
C_{\text{chain}}:=\left(\dfrac{\mathrm{d} \langle t \rangle_i}{\mathrm{d} d}\right)^{-1}= \alpha +\beta \,,
\label{eqn:chain_speed}
\end{align}
which agrees with the speed obtained from the observed arrival times.

Similarly, all measures of the response strength scale identically with increasing distance. Fig.~\ref{fig:directedchain}(d) illustrates this scaling for the first nodes in the chain and shows that the typical response magnitude $H_i$ and the response amplitude $x_i(\tpeak_i)$ differ by a constant factor, when the distance is large.

Together, this shows that the proposed measures accurately characterize the response strength and time. 

\subsection{Consistent quantification across topologies and perturbed units}

Can these measures also characterize the response for complex coupling topologies? To investigate how consistent the quantifiers are across different network topologies and the choice of the perturbed unit, we study 100 systems (Def.~\ref{dfn:linearsystem}) with random topologies, for each topology perturbing each unit one by one and measuring the responses. Fig.~\ref{fig:random_network} shows that, no matter which node is perturbed initially, the response is accurately characterized by the typical response time and magnitude. As for the chain, the typical response time $\left<t\right>_i$ is slightly larger than the peak response time $\tpeak_i$ by a constant and the typical response magnitude $H_i$ is larger than the response amplitude $x_i(\tpeak_i)$ by a constant factor. Across all 100 random topologies tested and all units perturbed in each network, the quantifiers 
consistently indicate the same distance-dependent response times and response magnitudes with only small deviations.

Results for other interaction topologies are qualitatively the same as shown in the appendix (see section \ref{sec:randomNetworksApp}).

In all figures we compare our measures to the time and height of the first peak of $x_i(t)$. Given a general network with general coupling strengths it is possible that $x_i(t)$ is not in fact unimodal and may have multiple peaks, especially in scale-free or star-like networks. Still, measures such as the total response strength and the expected arrival time describe the dynamics of the perturbation with only a few numbers. If needed, one can easily extend the above definitions to include higher-order measures such as skewness to more accurately describe the perturbation.

\section{Summary and Conclusions}

How networked systems transiently respond to external signals fundamentally underlies their robustness against perturbations. For instance, for a range of transient phenomena, such as the spreading of perturbations in power grids \cite{UCTE07_emsRiverReport} or of viral infections during an epidemic \cite{hufnagel04_forecast}, the transients are relevant because brief deviations may cause system-wide failures or undesired states, from overloads of transmission lines to power outages and from an increased number of infected individuals to secondary outbreaks. Yet dynamical systems theory so far mainly focuses on steady states, thereby neglecting the dynamics during transients. In particular it remained unclear how to analytically quantify arrival time or impact of a perturbation in a setting of deterministic network dynamical systems.

In this article, we changed the perspective on how to interpret transient trajectories and proposed quantitative measures for the arrival times and response strengths to perturbations, revealing how suitably normalized trajectories $x_i(t)$ may be viewed as equivalent to probability densities. This enabled us to intuitively define measures for the arrival times as the effective expectation values $\left<t\right>_i$ and the impact of the perturbation via the standard deviation $\sigma_i$ and resulting response magnitude $H_i$. 

We remark that already the total response $Z_i$ is a valid measure of response strength. However, which quantity is appropriate will depend on the system and the question about the transients to be answered. For instance, if a voltage excursion in an electric signal may not exceed a certain maximum to protect an electric device from shutdown, the \emph{magnitude of a response} $H_i$ might be a relevant quantity, whereas if a current charges a device that cannot safely store more than a certain amount of electrical charge, the \emph{total response} $Z_i$ is more suitable. In alternative settings, for instance, the total response strength $Z_i$ is clearly valuable describing the total number of infected individuals at a given location in models of epidemics and the expected value $\left<t\right>_i$ and standard deviation $\sigma_i$ provide key information about its arrival time and duration, indicating when the outbreak is most severe. Such quantities may provide valuable information across systems and, e.g., help suggesting periods and locations where additional preventive measures should be taken. 
 
For basic linear dynamical systems,  we derived simple analytical expressions writing all of those measures as direct functions of the inverse of the effective coupling matrix. We demonstrated that these \emph{expected value quantifiers} accurately describe the spreading of the perturbation across different network topologies and system parameters. They scale with distance in the same way as standard measures such as the time and height of the largest perturbation. As such, these measures provide an efficient analytical tool to predict and study transient spreading dynamics.

Finally, we remark that these expected value quantities can in principle be applied also to more general linear systems, independent even of the positivity of the trajectory, and may serve as qualitative and sometimes quantitative evaluators for nonlinear systems as well. For instance, in systems where a perturbation causes damped oscillations with alternating positive and negative periods of the state variables $x_i(t)$ or observable $g(x_i(t))$, computing the average  $\left<t\right>_i$ would often still provide a reasonable estimator for the order of magnitude of the typical response time. Moreover, nonlinear systems where the nonlinearities do not alter the qualitative form of the trajectory substantially until after the signal variation has almost ended, may be equally evaluated, because of the only minor influence of the tail of $\rho_i(t)$ on the effective expected values. How to extend the concept of expected value quantifiers to reveal further information about transient dynamics and how to generalize some of them to broader classes of linear and nonlinear systems needs to be explored in future research.\\

 \textit{Acknowledgments.} We thank L. Bunimovich, E. Estrada, Y. Fender, S. Grosskinsky, J. Nagler and N. Sharafi for valuable discussions and P. Ashwin for inspiring thoughts about transients in dynamical systems. This work was partially supported by the German Science Foundation (DFG) through the Center of Excellence `Center for Advancing Electronics Dresden' (cfaed). We gratefully acknowledge support from the Federal Ministry of Education and Research (BMBF Grant No. 03SF0472E). MT acknowledges hospitality by and a Fellowship as participant of an Advanced Study Group at the Max Planck Institute for the Physics of Complex Systems.

\bibliography{PerturbationSpreading}

\onecolumngrid

\newpage

\appendix

\section{Alternative derivation of the analytical expressions}
\label{app:proof}
In this section we give an alternative proof for expression for the total response strength $Z_i$ given by Eq.~(\ref{eqn:partitionFunction_matrix}) via integration of the linear differential equation defined in Eq.~(\ref{eqn:linearsystem}). Therefore we start by integrating Eq.~(\ref{eqn:linearsystem}) which gives
\begin{align}
\mathbf{x}(t)=M\mathbf{X}(t),
\end{align}  
where we substituted the definition of the indefinite integral
\begin{align}
\mathbf{X}(t) = \int \mathbf{x}(t) dt. 
\end{align}
We obtain
\begin{align}
\mathbf{X}(t) = M^{-1} \mathbf{x}(t)
\end{align}
which we substitute into the definition of the total response strength $Z_i$ from Eq. (\ref{eqn:partitionFunction})  
\begin{align}
Z_i&= \int_0^{\infty} x_i(t) dt\\ 
&= [X_i(t)]_0^{\infty}\\
&=[\left(M^{-1}  \mathbf{x}(t)\right)_i]_0^{\infty}\\
&= -\left(M^{-1} \mathbf{x}_0\right)_i
\end{align}
so that we obtain the same expression for the total response strength as derived in Eq.~(\ref{eqn:partitionFunction_matrix}).

\section{Network response measures in homogeneous chains}
\label{sec:homogeneousChainApp}

In this section we show the detailed calculation of the network response measures: the total response strength $Z_i$, the effective probability density $p_i(t)$, the typical response time $\langle t \rangle_i$, the effective standard deviation $\sigma_i$, and the response height $H_i$.

\subsection{Total repsonse strength $Z_i$ and effective probability density $\rho_i(t)$}
We start with the total response strength $Z_i$, which allows the calculation of the response density [Eq.~(\ref{eqn:probability}] and the typical repsonse time [Eq.~(\ref{eqn:meantime}]. To solve the integral in the definition of $Z_i$ [Eq.~(\ref{eqn:partitionFunction}], we have to find the indefinite integral
\begin{align}\label{chain_antiderivative_condition.eq}
X_i(t) := \int x_i(t)~dt.
\end{align}
We assume an ansatz
\begin{align}\label{chain_antiderivative.eq}
X_i(t) =- \frac{e^{-\beta t}}{\beta}- e^{-(\alpha + \beta) t}  \sum_{j=0}^{i-2} A_j \, t^j,
\end{align}
where the coefficients $A_j$ yet need to be determined. Calculating the time derivative of both sides of Eq.~(\ref{chain_antiderivative.eq}) yields the condition $\dot{X}_i(t) \overset{!}{=} x_i(t)$ given by the definition of $X_i(t)$ [Eq.~(\ref{chain_antiderivative_condition.eq})] and thereby the coefficients $A_j$. The time derivative of the right hand side of Eq.~(\ref{chain_antiderivative.eq}) reads
\begin{align}\label{chain_ansatz.eq}
\dot{X}_i(t) &= e^{-\beta t} - e^{-(\alpha + \beta) t}  \left( \sum_{j=0}^{i-2} j A_j t^{j-1} - (\alpha + \beta)  \sum_{j=0}^{i-2} A_j  t^j   \right).
\end{align}
Defining a new index $j'=j+1$ gives
\begin{align}\label{executesum.eq}
\dot{X}_i(t)&=e^{-\beta t} - e^{-(\alpha + \beta) t}  \left( \sum_{j=0}^{i-2} j A_j t^{j-1} - (\alpha + \beta)  \sum_{j'=1}^{i-1} A_{j'-1}  t^{j'-1}  \right).
\end{align}
Now we combine the two sums into one running from $1$ to $i-2$ and obtain
\begin{align}
\dot{X}_i(t)&= e^{-\beta t} - e^{-(\alpha +\beta) t} \left(\underbrace{-(\alpha + \beta)  A_{i-2}}_\text{$\overset{!}{=}\dfrac{\alpha^{i-2}}{(i-2)!}$ ~~ (\textrm{I})} \, t^{i-2} +\sum_{j=1}^{i-2} \underbrace{(j\, A_j - (\alpha + \beta) A_{j-1})}_\text{$\overset{!}{=}\dfrac{\alpha^{j-1}}{(j-1)!}$~~ (\textrm{II})} \, t^{j-1}  \right),
\end{align}
where the relations (\textrm{I}) and (\textrm{II}) follow from the condition that $\dot{X}_i(t) \overset{!}{=} x_i(t)$ with the solutions $x_i(t)$ in Eq.~(\ref{eqn:chain_solution}) 
\begin{align}
\dot{X}_i(t)\overset{!}=  e^{-\beta t}  \left(1 - e^{-\alpha t}  \sum_{j=0}^{i-2} \dfrac{(\alpha t)^j}{j!} \right)
\end{align}
for $i \geq 2$.
From relation (I) we obtain
\begin{align}
A_{i-2}&=-\dfrac{\alpha^{i-2}}{(i-2)!}  \dfrac{1}{(\alpha + \beta)},
\end{align}
and the remaining coefficients are obtained recursively using relation (\textrm{II}). This procedure yields the coefficients 
\begin{align}
A_j = -\sum_{\ell=j}^{i-2} \frac{\alpha^{\ell}}{j!  (\alpha +\beta)^{\ell-j+1}},\label{chain_A.eq}
\end{align}
for $j \in \{ 0, \ldots, i-2 \}$.
We use the indefinite intergral $X_i(t)$ we determined [Eq.~(\ref{chain_A.eq}) and Eq.~(\ref{chain_antiderivative.eq})] to calculate the total repsonse strength by its definition given in Eq.(\ref{eqn:partitionFunction})
\begin{align}
Z_i &= X_i(\infty) - X_i(0)\notag\\
&= \lim\limits_{t \to \infty} \left( - \frac{e^{-\beta t}}{\beta} - e^{-(\alpha + \beta) t}  \sum_{j=0}^{i-2} A_j  t^j \right) - \left( -\frac{1}{\beta} - A_0 \right).
\end{align}
The first term converges to zero, since $\displaystyle\lim_{t\rightarrow \infty}e^{-ct}=0$ for any $c\in \mathbb{R}_{\geq0}$. So does the second term because
\begin{align}\label{hospital.eq}
\lim\limits_{t \to \infty} e^{-ct} \sum_{i=0}^{n}  t^i =\lim\limits_{t\to\infty} \dfrac{\sum_{i=0}^{n}  t^i}{e^{ct}}\underset{\text{Rule}}{\overset{\text{L'Hospital's}}{=}}\lim\limits_{t \to \infty} \frac{n!}{c^n e^{ct}} = 0
\end{align}
holds for finite $n \in \mathbb{N}$. 
Thus we have
\begin{align}
Z_i&\overset{\phantom{\text{Eq.(\ref{chain_A.eq})}}}{=} \frac{1}{\beta} + A_0\notag\\
&\overset{\text{Eq.(\ref{chain_A.eq})}}{=} \frac{1}{\beta} - \sum_{j=0}^{m-2} \frac{\alpha^j}{(\alpha +\beta)^{j+1}}\notag\\
&\overset{\phantom{\text{Eq.(\ref{chain_A.eq})}}}=\frac{1}{\beta} -\frac{1}{\alpha +\beta}\sum_{j=0}^{i-2} \frac{\alpha^j}{(\alpha +\beta)^{j}}\notag\\
&\overset{\phantom{\text{Eq.(\ref{chain_A.eq})}}}= \frac{1}{\beta} -\frac{1}{\alpha +\beta} \left(\frac{1-{\left(\frac{\alpha}{\alpha +\beta}\right)}^{i-1}}{1-\frac{\alpha}{\alpha +\beta}}\right)
\end{align}
for all units $i \geq 2$. For $i=1$ using the solution given in Eq.~(\ref{eqn:chain_solution}) the total response strength becomes
\begin{align}
Z_1=\frac{1}{\beta}.
\end{align}
With further simplifications we obtain
\begin{align}
\label{chain_Z.eq}
\boxed{
Z_i=\dfrac{1}{\beta}\left({\dfrac{\alpha}{\alpha +\beta}}\right)^{i-1}}
\end{align}
for all $i \in \{ 1, \ldots, N \}$.

Now we calculate the effective probability density directly following the definition [Eq.~(\ref{eqn:probability})]. Using the expression of the total repsonse strength [Eq.~(\ref{chain_Z.eq})], we obtain
\begin{align}
\rho_i(t)=
\left({\frac{\alpha}{\alpha +\beta}}\right)^{1-i}\beta  e^{-\beta t}  \left(1 - e^{-\alpha t}  \sum\limits_{j=0}^{i-2} \dfrac{(\alpha t)^j}{j!} \right).
\end{align}

\subsection{Typical response time $\langle t \rangle_i$}
As the next step we calculate the typical response time $\langle t \rangle_i$ based on the previous results. Following the definition of $\langle t \rangle_i$ in Eq. (\ref{eqn:meantime}) and the definition of the effective probability density $\rho_i(t)$ in Eq. (\ref{eqn:probability}), we have by partial integration
\begin{align}
\langle t\rangle_i&=\dfrac{1}{Z_i}  \left(X_i(t)  t ~\Big|_{t=0}^{t\rightarrow\infty}- \int_0^{\infty} X_i(t') ~dt' \right),
\end{align}
where $X_i(t)$ is defined in Eq.~(\ref{chain_antiderivative_condition.eq}) and given by Eq.~(\ref{chain_antiderivative.eq}) and Eq.~(\ref{chain_A.eq}). Substituting $Z_i$ with the expression in Eq.~(\ref{chain_antiderivative.eq}) yields
\begin{align}
\langle t\rangle_i = \frac{1}{Z_i} \left(t \left(\dfrac{-e^{-\beta t}}{\beta} - e^{-(\alpha + \beta) t}  \sum_{j=0}^{i-2} A_i  t^i \right) ~\Big|_{t=0}^{t\rightarrow\infty} \right. \left.+ \int_0^{\infty}  \left(\frac{e^{-\beta t'}}{\beta} + e^{-(\alpha + \beta) t'}  \sum_{i=0}^{m-2} A_i  t'^i \right) dt' \right).
\end{align}
It is easy to see that the first term vanishes [cf. Eq.~(\ref{hospital.eq})], hence the expression becomes
\begin{align}
\langle t\rangle_i &=\frac{1}{Z_i} \left( \int_0^{\infty}  \frac{e^{-\beta t'}}{\beta} ~~dt' + \int_0^{\infty}  \left(e^{-(\alpha + \beta) t'}  \sum_{j=0}^{i-2} A_j  t'^j \right) dt' \right). 
\end{align}
The first integral is easy to solve:
\begin{align}\label{chain_intX.eq}
\int_0^{\infty}  \frac{e^{-\beta t}}{\beta} ~dt=\frac{-e^{-\beta t}}{{\beta}^2}\Bigr\rvert_0^{\infty}=\frac{1}{{\beta}^2},
\end{align}
whereas the second one can be solved with a similar method as used for calculating the total response strength $Z_i$. We define 
\begin{align}\label{chain_defF.eq}
F_i(t):=\int_{-\infty}^t \left( e^{-(\alpha + \beta) t'}  \sum_{j=0}^{i-2} A_j  t'^j \right)dt' 
\end{align}
so that the typical response time $\langle t\rangle_i$ can be written as
\begin{align}
\langle t\rangle_i &= \dfrac{1}{Z_i} \left( \dfrac{1}{{\beta}^2} + F_i(\infty) - F_i(0) \right). \label{chain_integral_solution.eq}
\end{align}
Again we assume an ansatz for the integral
\begin{align}
F_i(t) &= e^{-(\alpha + \beta) t}  \sum_{j=0}^{i-2} B_j \, t^j. \label{chain_ansatzF.eq}
\end{align}
According to the definition [Eq.~(\ref{chain_defF.eq})], the time derivative of $F_i(t)$ has to obey
\begin{align}
\dot{F}_i(t) \overset{!}{=} e^{-(\alpha + \beta) t}  \sum_{j=0}^{i-2} A_j \, t^j. \label{chain_conditionF.eq} 
\end{align}
Inserting the ansatz for $F_i(t)$ [Eq.~(\ref{chain_ansatzF.eq})] into Eq.~(\ref{chain_conditionF.eq}) and comparing the coefficients allows to determine the coefficients $B_j$. We take the time derivative of the ansatz and obtain
\begin{align}\label{chain_methodstart.eq}
\dot{F}_i(t) &= e^{-(\alpha + \beta) t}  \left( \sum_{j=0}^{i-2}i  B_j  t^{j-1} - (\alpha + \beta)  \sum_{j=0}^{i-2} B_j  t^j \right).
\end{align}
Defining a new index $j'=j+1$ for the second sum to shift the order of $t$ yields
\begin{align}
\dot{F}_i(t)&= e^{-(\alpha + \beta) t}  \left( \sum_{j=0}^{i-2}j  B_j  t^{j-1} - (\alpha + \beta)  \sum_{j'=1}^{i-1} B_{j'-1}  t^{j'-1} \right).
\end{align}
Again we combine the sums into one and compare the coefficients, thus obtain
\begin{align}
\dot{F}_i(t)&= e^{-(\alpha + \beta) t}  \left(\underbrace{-(\alpha + \beta)  B_{i-2}}_\text{$\overset{!}{=}A_{i-2}$} t^{i-2} + \sum_{j=1}^{i-2} \underbrace{(j  B_j - (\alpha + \beta)  B_{j-1})}_\text{$\overset{!}{=}A_{j-1}$}  t^{j-1}\right).
\end{align}
The cofficients of the highest order of $t$ reads
\begin{align}
B_{i-2}= \dfrac{-A_{i-2}}{(\alpha + \beta)}.
\end{align}
The remaining coefficients are again obtained recursively
\begin{align}
B_j = \frac{(j+1)  B_{j+1}}{(\alpha + \beta)}- \frac{A_j}{(\alpha + \beta)}.
\end{align}
Hence, the general expression of coefficients can be written as 
\begin{align}\label{chain_B.eq}
B_j = - \sum_{\ell=j}^{i-2} \dfrac{A_{\ell}}{{(\alpha + \beta)}^{\ell-j+1}} \frac{\ell !}{j!}
\end{align} 
for $j \in \lbrace 0,\cdots,i-2\rbrace$.
Now we calculate the typical response time $\langle t \rangle_i$ using the expression of $F_i(t)$ given by Eq.~(\ref{chain_ansatzF.eq}) and Eq.~(\ref{chain_B.eq}). Writing $F_i(t)$ explicitly in Eq.~(\ref{chain_integral_solution.eq}) yields
\begin{align}
\langle t \rangle_i &= \frac{1}{Z_i} \left(\frac{1}{{\beta}^2}+\lim\limits_{t \to \infty} e^{-(\alpha + \beta) t}  \sum_{j=0}^{i-2} B_j  t^j - B_0 \right).
\end{align}
As discussed [cf. Eq.~(\ref{hospital.eq})], the term in the middle converges to zero, which leaves
\begin{align}
\langle t \rangle_i&= \frac{1}{Z_i} \left(\frac{1}{{\beta}^2} - B_0 \right).\label{chain_meant.eq}
\end{align}
Here $B_0$ can be determined using Eq.~(\ref{chain_B.eq}) and Eq.~(\ref{chain_A.eq}) as
\begin{align}
B_0&= - \sum_{j=0}^{i-2} \frac{A_j j!}{{(\alpha + \beta)}^{j+1}}\notag\\
&= \sum_{j=0}^{i-2} \frac{j!}{{(\alpha + \beta)}^{j+1}} \sum_{k=j}^{i-2} \frac{\alpha^k}{j!  (\alpha +\beta)^{k-j+1}}\notag \\
&=\sum_{j=0}^{i-2} \frac{1}{{(\alpha + \beta)}^{2}} \sum_{k=j}^{i-2} \left(\frac{\alpha}{\alpha +\beta}\right)^{k}.
\end{align}
Noticing that the sums are determined by the formula for summing geometric series, we further simplify the expression of $B_0$ and obtain
\begin{align}
B_0= \frac{1}{{\beta}^2}-\frac{\left(\frac{\alpha}{\alpha +\beta}\right)^{i-1}}{(\alpha + \beta)} \left(\frac{\alpha+i \beta}{{\beta}^2} \right).
\end{align}

Using this result and the expression of the total response strength (Eq.~\ref{chain_Z.eq}), we thus obtain the effective response time
\begin{align}
\label{chain_meantsolution.eq}
\boxed{
\langle t \rangle_i= \frac{\alpha + i \beta}{\alpha \beta + {\beta}^2}.}
\end{align}
We notice that $\langle t \rangle_i$ shows a linear dependence on the index of unit $i$:
\begin{align}\label{chain_tmeanslope.eq}
\frac{d \langle t \rangle_i }{di} =\frac{1}{\alpha+\beta}.
\end{align}
That means, in a homogeneous directed chain, the perturbation spreads with a constant speed $\frac{1}{\alpha+\beta}$, if we measure the arrival time of the perturbation with the effective response time $\langle t \rangle_i$.

\subsection{Typical response duration ${\sigma}_{i}$ and response magnitude $H_i$}
Next we derive the effective standard deviation $\sigma_i$  which we interpret as the typical response duration [Eq. (\ref{eqn:duration})] and the typical response magnitude $H_i$ [Eq. (\ref{eqn:height})], which quantify the width and the height of the response profile. First we calculate the second central moment of $t$ and using the result the square of the effective standard deviation [Eq.~(\ref{eqn:duration})]. The second moment of $t$ is given as
\begin{align}
\langle t^2 \rangle_i = \int_0^{\infty} \rho_i(t') t'^2 ~dt'
\overset{\text{Eq.\ref{eqn:probability}}}{=} \frac{1}{Z_i}\int_0^{\infty} x_i(t') t'^2 ~dt'.
\end{align}
Partial integration yields
\begin{align}\label{chain_2moment.eq}
Z_i\langle t^2 \rangle_i = X_i(t) t^2\big|_0^{\infty} -2\int_0^{\infty} X_i(t') t' ~dt',
\end{align}
where $X_i(t)$ is defined above in Eq.~(\ref{chain_antiderivative_condition.eq}).
To determine the integral in Eq.~(\ref{chain_2moment.eq}), we define 
\begin{align}\label{chain_Ftilde.eq}
\tilde{F}_i(t):= \int_{-\infty}^t X_i(t')~dt'.
\end{align}
Using the expression of $X_i(t)$ [Eq.~(\ref{chain_antiderivative.eq})] and $F_i(t)$ [Eq.~(\ref{chain_defF.eq})], we obtain the following relation between $\tilde{F}_i(t)$ and $F_i(t)$:
\begin{align}
\tilde{F}_i(t)&\overset{\text{Eq.\ref{chain_antiderivative.eq}}}{=}\int_{-\infty}^t \left(- \frac{e^{-\beta t'}}{\beta}- e^{-(\alpha + \beta) t'}  \sum_{j=0}^{i-2} A_j \, t'^j\right) ~dt' \notag\\
&\overset{\text{Eq.\ref{chain_defF.eq}}}= \int_{-\infty}^t \left(- \frac{e^{-\beta t'}}{\beta}\right)~dt'- F_i(t)\notag\\
&\overset{\text{Eq.\ref{chain_intX.eq}}}= \dfrac{e^{-\beta t}}{{\beta}^2} - F_i(t).\label{chain_FtildeF.eq}
\end{align}
Expressing the integral in Eq.~(\ref{chain_2moment.eq}) in terms of $\tilde{F}_i(t)$ and using partial integration again, we have
\begin{align}
Z_i\langle t^2 \rangle_i &=  X_i t^2\bigr\rvert_0^{\infty} -2 \left(\tilde{F}_i(t) t \bigr\rvert_0^{\infty} - \int_0^{\infty} \tilde{F}_i(t') dt'\right)\notag\\
&= \underbrace{X_i t^2\bigr\rvert_0^{\infty}-2 \tilde{F}_i(t) t \bigr\rvert_0^{\infty}}_{=0}+2 \int_0^{\infty} \tilde{F}_i(t') dt'.
\end{align}
The first two terms vanish each according to Eq.~(\ref{hospital.eq}). We then use the relation Eq.~(\ref{chain_FtildeF.eq}), thus
\begin{align}\label{chain_intFbar.eq}
Z_i \langle t^2 \rangle_i&= 2\int_0^{\infty} \frac{e^{-\beta t'}}{{\beta}^2} ~dt' - 2 \int_0^{\infty} F_m(t') dt'\notag\\
&= -\left.\dfrac{2e^{-\beta t}}{{\beta}^3}\right|_0^{\infty}  - 2 \int_0^{\infty} F_m(t') dt'.
\end{align}
By futher defining
\begin{align}
\bar{F}_i(t) := \int_{-\infty}^t F_i(t') ~dt',
\end{align}
we write $Z_i \langle t^2 \rangle_i$ in terms of the integral $\bar{F}_i(t)$:
\begin{align}
Z_i \langle t^2 \rangle_i&= \frac{2}{{\beta}^3}-2 \bar{F}_i(t) \bigr\rvert_0^{\infty}=\frac{2}{{\beta}^3}-2 \bar{F}_i( \infty) +2 \bar{F}_i(0)
\end{align}
In analogy to the method for deriving $F_i(t)$, we again assume an ansatz for $\bar{F}_i(t)$
\begin{align}\label{chain_Fbar.eq}
\bar{F}_i(t) = e^{-(\alpha +\beta)t} \sum_{j=0}^{i-2} C_j t^j,
\end{align}
which by definition obeys
\begin{align}
\dot{\bar{F}}_i(t) \overset{!}= F_i(t)\overset{\text{Eq.\ref{chain_ansatzF.eq}}}= e^{-(\alpha + \beta) t}  \sum_{j=0}^{i-2} B_j \, t^j. 
\end{align}
Again, by taking the derivative of the ansatz of $\bar{F}_i(t)$ and comparing the coefficients, as we did before in deriving $F_i(t)$ [Eq.~(\ref{chain_methodstart.eq} - \ref{chain_B.eq})], we obtain the coefficients
\begin{align}\label{chain_C.eq}
C_j &= - \sum_{\ell=j}^{i-2} \frac{B_j}{{(\alpha + \beta)}^{\ell-j+1}} \frac{\ell !}{j!}.
\end{align}
Now we determine $Z_i \langle t^2 \rangle_i$ by means of the expression of $\bar{F}_i(t)$. Inserting Eq.~(\ref{chain_Fbar.eq}) into Eq.~(\ref{chain_intFbar.eq}), we have 
\begin{align}\label{chain_intX2.eq}
Z_i\langle t^2 \rangle_i &= \frac{2}{{\beta}^3}-0 +2 C_0\notag\\
&=\frac{2}{{\beta}^3}+2 C_0.
\end{align}
Using the expression of $B_j$ [Eq.~(\ref{chain_B.eq})] and $A_j$ [Eq.~(\ref{chain_A.eq})], we determine $C_0$ as follows:
\begin{align}
C_0&=- \sum_{j=0}^{i-2} \frac{B_j}{{(\alpha + \beta)}^{j+1}} j!\notag\\
&= \sum_{j=0}^{i-2} \frac{j!}{{(\alpha + \beta)}^{j+1}}  \sum_{k=j}^{i-2} \frac{k!}{j!}\frac{1}{{(\alpha + \beta)}^{k-j+1}} \sum_{\ell=k}^{i-2} \frac{{\alpha}^{\ell}}{{(\alpha + \beta)}^{\ell-k+1}} \frac{1}{k!}\notag\\
&=  \frac{1}{{(\alpha + \beta)}^3} \sum_{j=0}^{i-2} \sum_{k=j}^{i-2}\sum_{\ell=k}^{i-2} \left(\frac{{\alpha}}{{\alpha + \beta}}\right)^{\ell} \notag\\
&=\frac{1}{{\beta}^3}-\left(\frac{\alpha}{\alpha+\beta}\right)^{i-1} \frac{1}{\beta}\left(\frac{1}{{\beta}^2} + \frac{(i-1)}{(\alpha+\beta)\beta}  + \frac{(i-1)}{(\alpha+\beta)^2}+\frac{(i-2)(i-1)}{2(\alpha+\beta)^2}\right).
\end{align}
In the last equation we executed the geometric sums.
Substituting $C_0$ and the total response strength $Z_i$ [Eq.~(\ref{chain_Z.eq})] into Eq.~(\ref{chain_intX2.eq}) gives
\begin{align}
\label{chain_meansquare.eq}
\boxed{
\langle t^2 \rangle_i= \frac{2{\alpha}^2+2\alpha \beta (i+1)+ i {\beta}^2 (i+1)}{(\alpha \beta + {\beta}^2)^2}.}
\end{align}
Now we obtain the typical response duration $\sigma_i$ using its definition [Eq.~(\ref{eqn:duration})], the expression of the typical response time $\langle t \rangle_i$ [Eq.~(\ref{chain_meantsolution.eq})] and the second moment $\langle t^2 \rangle_i$ [Eq.~(\ref{chain_meansquare.eq})]:
\begin{align}
\label{chain_standdev.eq}
\boxed{
\sigma_i= \dfrac{\sqrt{\alpha^2+2\alpha \beta+ i \beta^2}}{(\alpha \beta + \beta^2)}.}
\end{align}
Hence, by definition [Eq.~(\ref{eqn:height})], the typical response magnitude $H_i$ is given by
\begin{equation}
\label{chain_H.eq}
\boxed{
H_i=\frac{(\alpha+\beta)}{\sqrt{{\alpha}^2+2\alpha \beta+i {\beta}^2}}\left(\frac{\alpha}{\alpha+\beta}\right)^{i-1}.}
\end{equation}

\section{Quantifying spreading across network ensembles}
\label{sec:randomNetworksApp}

\begin{figure}[h!]
\centering
\includegraphics[width=130mm]{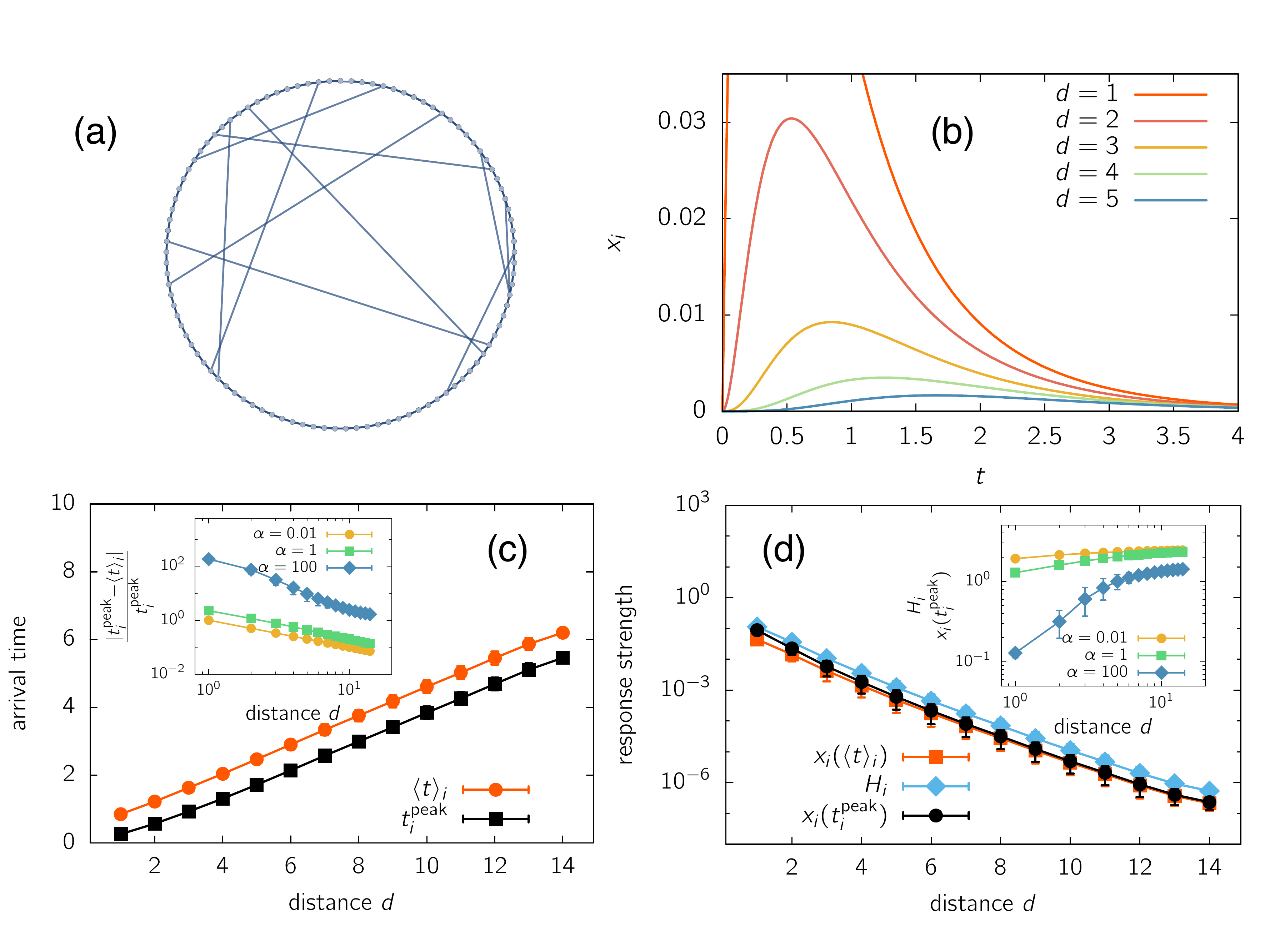}
\caption{ (color online)
\textbf{Perturbation spreading in a random small-world network.}
Illustration of the measures introduced above, describing approximate time and impact of the perturbation. (a) Illustration of the network topology, a small-world network \cite{watts98_smallworld} constructed from a ring of $N=100$ nodes. Each node is connected to its two nearest neighbors on either side (for a total of 4 connections) and $10$ links are uniformly randomly added to create shortcuts in the network.
As always, we choose $\beta = 1$. All results are averaged over $10$ different realizations of the network topology and perturbation of all nodes. Error bars indicate the standard deviation. (b) Example response dynamics of 5 nodes in the network with different distances to the initial perturbation. (c) Comparison of the typical response time $\left<t\right>_i$ and the peak response time $t_\mathrm{\mathrm{peak},i}$ for $\alpha = 1$. (d) Measurements of the strength of the perturbation given by the typical response strength $H_i$, the response at the typical response time $x_i\left(\left<t\right>_i\right)$ and the  response amplitude $x_i\left(t_\mathrm{\mathrm{peak},i}\right)$.}
\label{fig:smallWorld}
\end{figure}

\begin{figure}
\centering
\includegraphics[width=130mm]{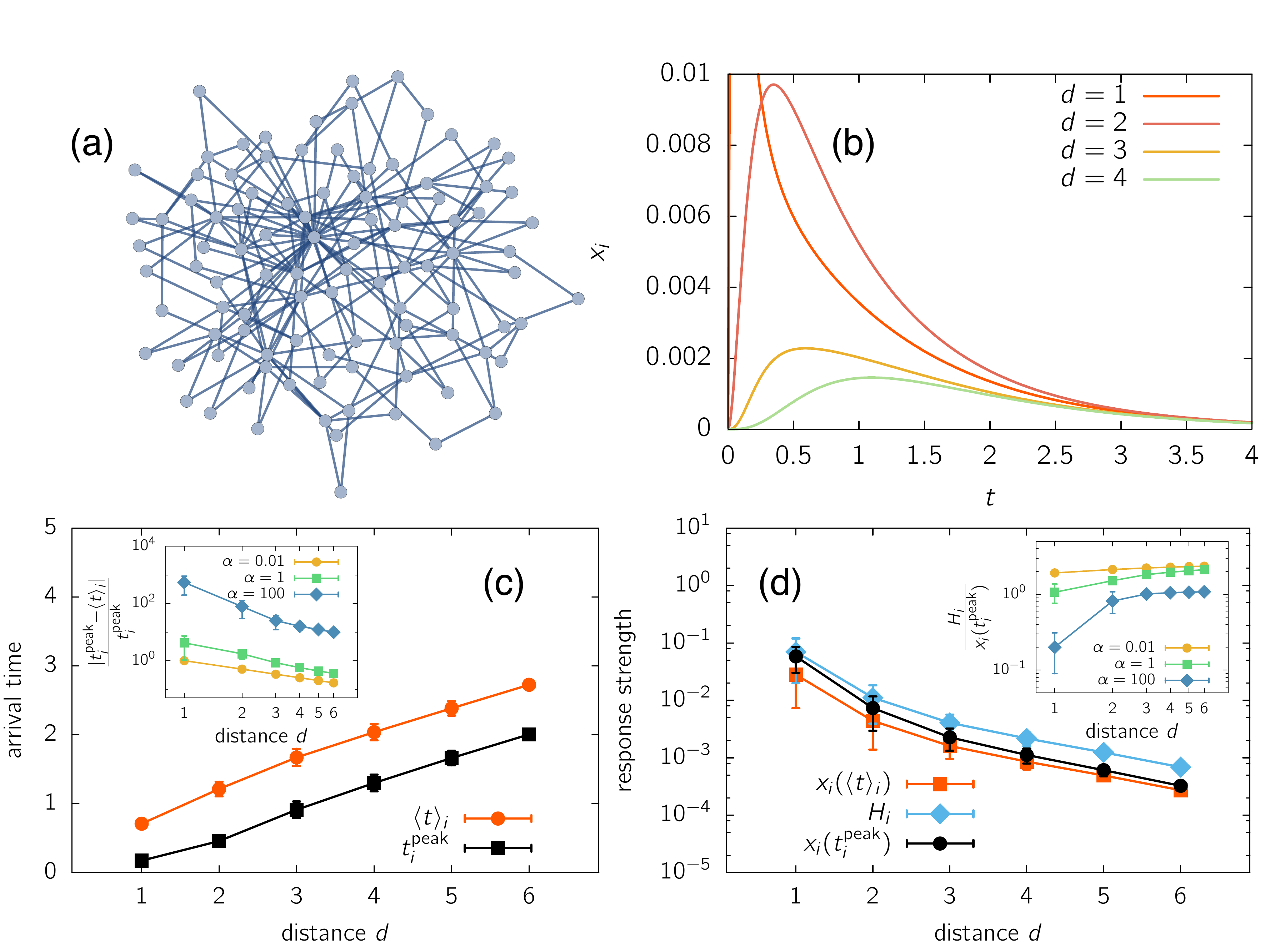}
\caption{ (color online)
\textbf{Perturbation spreading in a random scale-free network.}
Illustration of the measures introduced above, describing approximate time and impact of the perturbation. (a) Illustration of the network topology, a scale-free network with $N=100$ nodes. The network is constructed  by sequentially adding nodes with two links to the network, starting from a core of $5$ fully connected nodes. New nodes are attached to the network following the preferential attachment mechanism \cite{BAnetwork99}.
As always, we choose $\beta = 1$. All results are averaged over $10$ different realizations of the network topology and perturbation of all nodes. Error bars indicate the standard deviation. (b) Example response dynamics of 4 nodes in the network with different distances to the initial perturbation. (c) Comparison of the typical response time $\left<t\right>_i$ and the peak response time $t_\mathrm{\mathrm{peak},i}$ for $\alpha = 1$. (d) Measurements of the strength of the perturbation given by the typical response strength $H_i$, the response at the typical response time $x_i\left(\left<t\right>_i\right)$ and the  response amplitude $x_i\left(t_\mathrm{\mathrm{peak},i}\right)$.}
\label{fig:scaleFree}
\end{figure}

\begin{figure}
\centering
\includegraphics[width=130mm]{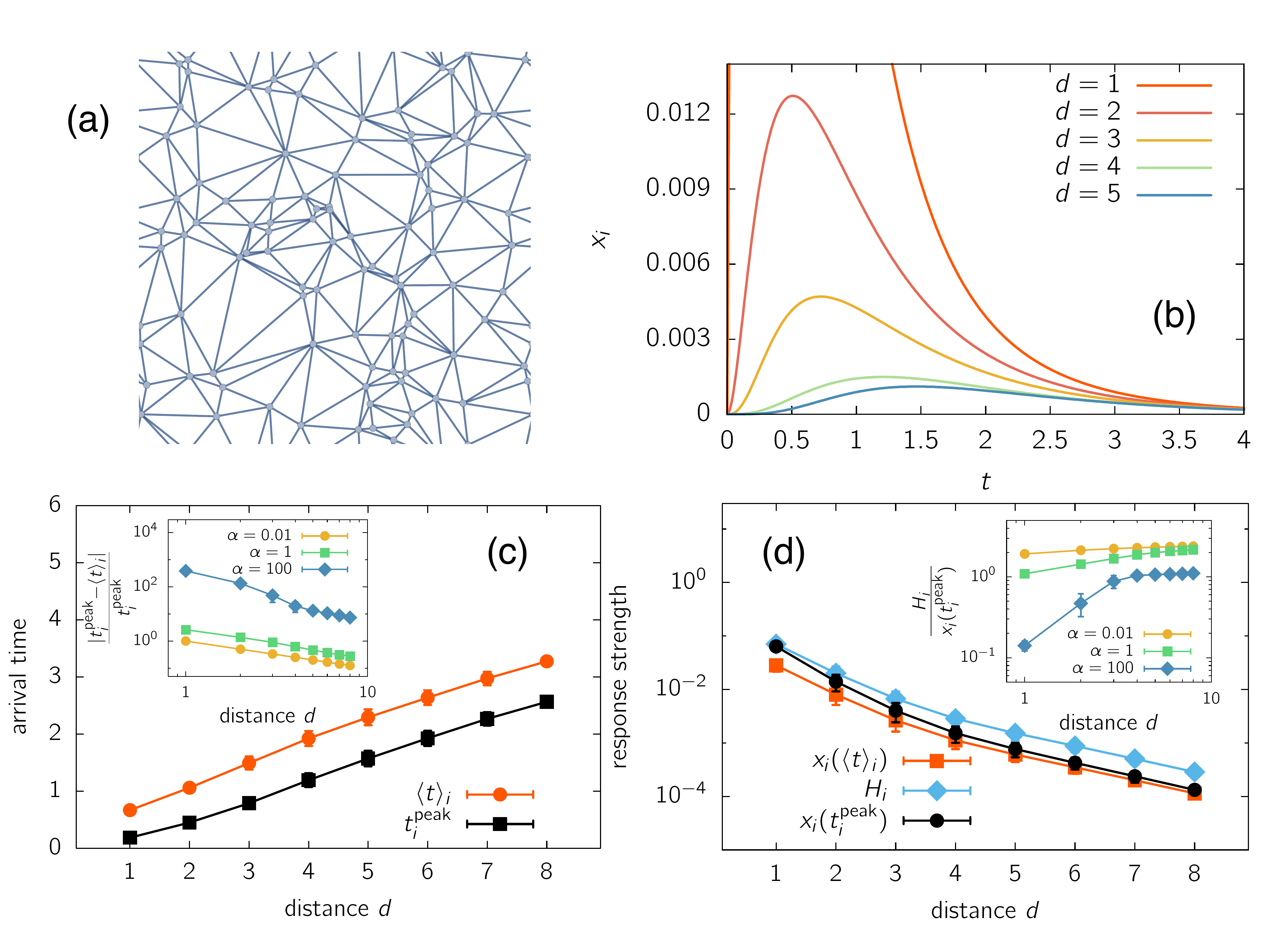}
\caption{ (color online)
\textbf{Perturbation spreading in a random geometric network.}
Illustration of the measures introduced above, describing approximate time and impact of the perturbation. (a) Illustration of the network topology, a random, geometrically embedded network with $N=100$ nodes. The network is constructed as a periodic Delaunay triangulation of $100$ points uniformly randomly distributed in the unit square.
As always, we choose $\beta = 1$. All results are averaged over $10$ different realizations of the network topology and perturbation of all nodes. Error bars indicate the standard deviation. (b) Example response dynamics of 5 nodes in the network with different distances to the initial perturbation. (c) Comparison of the typical response time $\left<t\right>_i$ and the peak response time $t_\mathrm{\mathrm{peak},i}$ for $\alpha = 1$. (d) Measurements of the strength of the perturbation given by the typical response strength $H_i$, the response at the typical response time $x_i\left(\left<t\right>_i\right)$ and the  response amplitude $x_i\left(t_\mathrm{\mathrm{peak},i}\right)$.}
\label{fig:geometric}
\end{figure}

In this section we give additional simulation results. We show that across various network topologies that the ``effective expectation values'' we proposed provide consistent measures of the response characteristics independent of the network topology.
We consider different network topologies and perform the same analysis on them as in Sec.~\ref{sec:examples}, averaging the result of a total of 1000 different initial perturbations. For small-world networks [Fig.~\ref{fig:smallWorld}], scale-free networks [Fig.~\ref{fig:scaleFree}] and random geometric networks [Fig.~\ref{fig:geometric}] we observe qualitatively the same results as obtained in Fig.~\ref{fig:directedchain}. Thus, also for differing network topologies the typical response time and the typical response magnitude give a consistent description of the response dynamics.

\vspace{-0.4cm}

\end{document}